\documentclass{iau} 
\usepackage{graphicx}

\title[AGB Variables in Nearby Galaxies] 
{Asymptotic Giant Branch Variables in Nearby Galaxies}

\author[Patricia A. Whitelock]   
{Patricia A. Whitelock}

\affiliation{ South African Astronomical Observatory, P O Box 9, 7935 Observatory, South Africa \&  Department of Astronomy, University of Cape Town, 7701 Rondebosch, South Africa\\email: {\tt paw@saao.ac.za}}

\pubyear{2018}
\volume{343}  
\jname{Why Galaxies Care About AGB Stars}
\editors{ Franz Kerschbaum, Martin Groenewegen \& Hans Olofsson, eds.}
\begin{document}

\maketitle

\begin{abstract}
Certain types of large amplitude AGB variable are proving to be powerful distance indicators that will rival Cepheids in the JWST era of high precision infrared photometry. These are predominantly found in old populations and have low mass progenitors. At the other end of the AGB mass-scale, large amplitude variables, particularly those undergoing hot bottom burning, are the most luminous representatives of their population. These stars are $<1$\,Gyr old, are often losing mass copiously and are vital to our understanding of the integrated light of distant galaxies as well as to chemical enrichment. However, the evolution of such very luminous AGB variables is rapid and remains poorly understood. Here I discuss recent infrared observations of both low- and intermediate-mass Mira variables in the Local Group and beyond.
\end{abstract}

\firstsection 
\section{Introduction}
Mira variables have the largest amplitudes ($\Delta V > 2$, $\Delta K >0.4$, $\Delta [4.5]> 0.3$ mag)  of any regularly pulsating star, which makes them relatively easy to identify. They pulsate at the fundamental frequency, as do a few semi-regular (SR) variables, although most of the SRs pulsate in one or more overtone modes (e.g., Trabucchi {\it et al.} 2017). 
In this presentation I want to address two topics which make Mira variables particularly interesting and important.  The first is Miras as distance indicators; there has been a lot of work in this area over the last few years and it becomes increasingly important as we move into the JWST era because Miras are such strong infrared sources. At the moment this involves exclusively short period ($P<400$ days) Miras. 

The second topic is the last stage in the AGB evolution of intermediate mass AGB stars. These long period stars, including new Miras that are being discovered in the Local Group and beyond, are not well understood. They can be carbon- or oxygen-rich; many of them have thick circumstellar shells and, depending on their initial metallicity, some are undergoing hot bottom burning. These long period stars are very important to our understanding of mass loss and provide fascinating probes of the late stages of stellar evolution. It is also at this stage of their evolution that the elusive massive, super-AGB, stars are most likely to be found.

In the context of the distance scale it is important to recognize that Miras are really big stars; their angular diameters are approximately twice their parallaxes. Furthermore, they have large convection cells so that their surface features are usually both asymmetric, and variable (e.g., Paladini {\it et al.} 2017). Thus Gaia is not easily going to give us distances. It may eventually be possible to disentangle the diameters and the surface features of some very well observed stars, but that sort of analysis will take time and will not be a part of early Gaia data releases. 
van Langevelde (2018) also discussed this point in the context of radio VLBI measures of Miras. VLBI parallaxes are very useful for Miras, but are time consuming to obtain and only applicable to O-rich stars which have OH, $\rm H_2O$, or SiO Masers.

\section{Miras as Distance Indicators}
The near-infrared period-luminosity relation (PLR) for the Miras in the LMC has been investigated over many years (Feast {\it et al.} 1989; Wood {\it et al.} 1999; Whitelock  {\it et al.} 2008;   Ita \& Matsunaga 2011; Yuan {\it et al.} 2017).  The details depend on the wavelength, but in general stars with thick dust shells will fall below the PLR because of circumstellar extinction and this is most obvious for C-rich stars and at shorter infrared wavelengths. If the extinction can be corrected these C-stars seem to fall on the same PLR as the O-rich ones.

\begin{figure}
\centering
\includegraphics[width=0.42\linewidth]{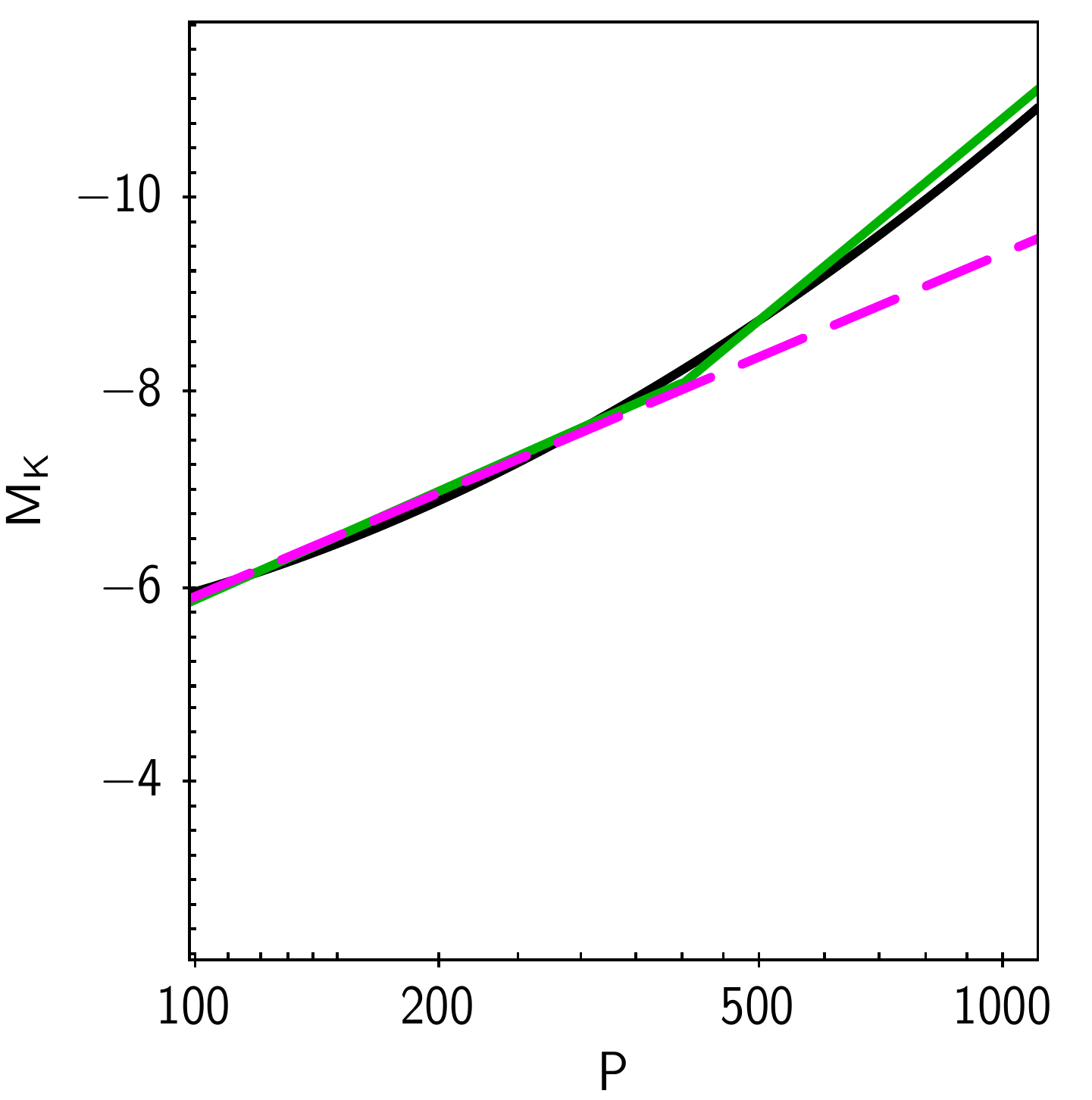}\includegraphics[width=0.45\linewidth]{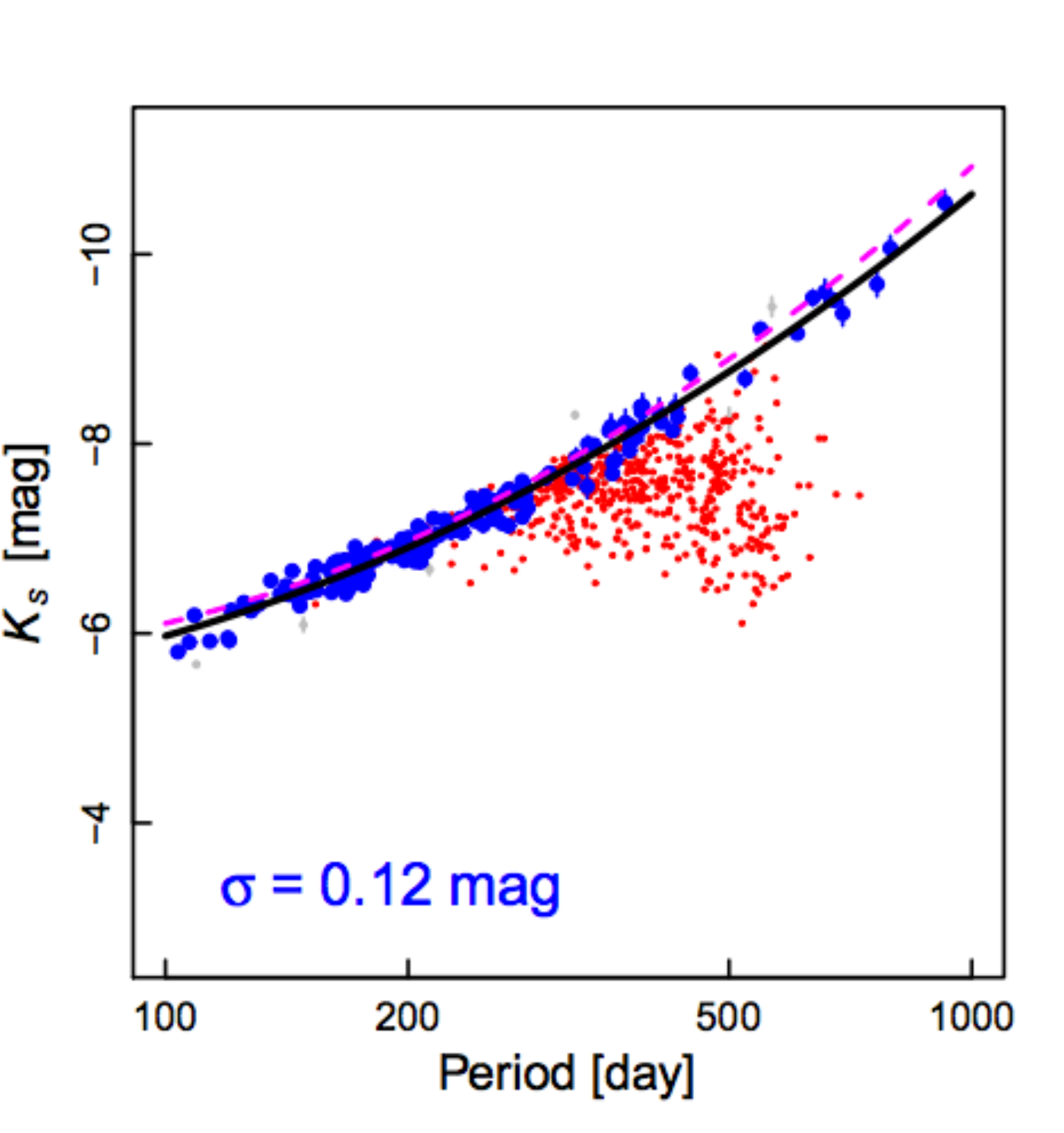}
\caption{(right) the Yuan {\it et al.} (2017) parabolic fit to the LMC O-rich Miras; O-rich and C-rich stars in blue and red, respectively; (left) a comparison of the PLRs from Whitelock {\it et al.} (2008)(magenta), Ita \& Matsunaga (2011)(green) and Yuan {\it et al.} (2017)(black).}
 \label{fig1}
\end{figure}

Figure~\ref{fig1} shows the parabolic $K_S$-PLR derived by Yuan {\it et al.} (2017)  for O-rich Miras in the LMC, using mean magnitudes, and compares it with the two linear PLRs derived earlier by Ita \& Matsunaga (2011), for more or less the same stars although using single observations,  and the extrapolated PLR derived for Galactic and LMC mean magnitudes by Whitelock {\it et al.} (2008).  For O-rich stars with $P<400$ days the three PLRs are essentially identical. For stars with longer periods the extrapolated Whitelock et al. relation is different from the other two (noting that none of the curves are well defined at $P>>500$ days). The Yuan {\it et al.}  approach nicely illustrates what can be achieved for these large amplitude variables using mean magnitudes. 

The details of pulsation in the fundamental mode have not yet been properly modeled (Trabucchi {\it et al.} 2017), but the linear PLR for short period Miras is presumably a consequence of the core-mass-luminosity relation (Paczynski 1970) very close to the end of the AGB phase in low mass stars. So the pulsation period is a function of the initial mass of these stars, as well as their current mass (e.g. Feast 2009).   

Discussions of the bolometric PLR indicated that C-rich stars fall on the same linear PLR as the short period Miras over a large range of periods as seen in, e.g. the dwarf irregulars NGC\,6822 (Whitelock {\it et al.} 2013) and IC\,1613 (Menzies {\it et al.} 2015). The bolometric magnitudes are calculated by applying a colour dependent bolometric correction to the $K$ magnitude\footnote{This does not necessarily provide the best possible measure of bolometric luminosity, but it is simple and can be applied to measurments from different galaxies to derive distances.}. The linear PLR can be understood if the C-rich stars obey the same core-mass-luminosity relation as the short period O-rich stars. What is lacking at this stage is a theoretical explanation of why most stars pulsate in the fundamental mode (i.e. become Miras) only at the very end of their AGB lifetimes and hence obey a linear PLR. 

Yuan {\it et al.} (2018) applied the method developed by He {\it  et al.} (2016) to Miras in M33 and used detailed fitting of the $I$-band light curve  to evaluate mean $JHK_S$ magnitudes from a small number of observations. They found that the C-stars lay below the PLR at $J$ and $H$ (as expected due to circumstellar extinction), but very close to it at $K_S$. The distance they derive is in reasonable agreement with the values found using Cepheids (Gieren {\it et al.} 2013 and references therein), RR Lyr variables (Sarajedini {\it et al.} 2006) and eclipsing binaries (Bonanos {\it et al.} 2006).

\begin{figure}
\centering
\includegraphics[height=5cm]{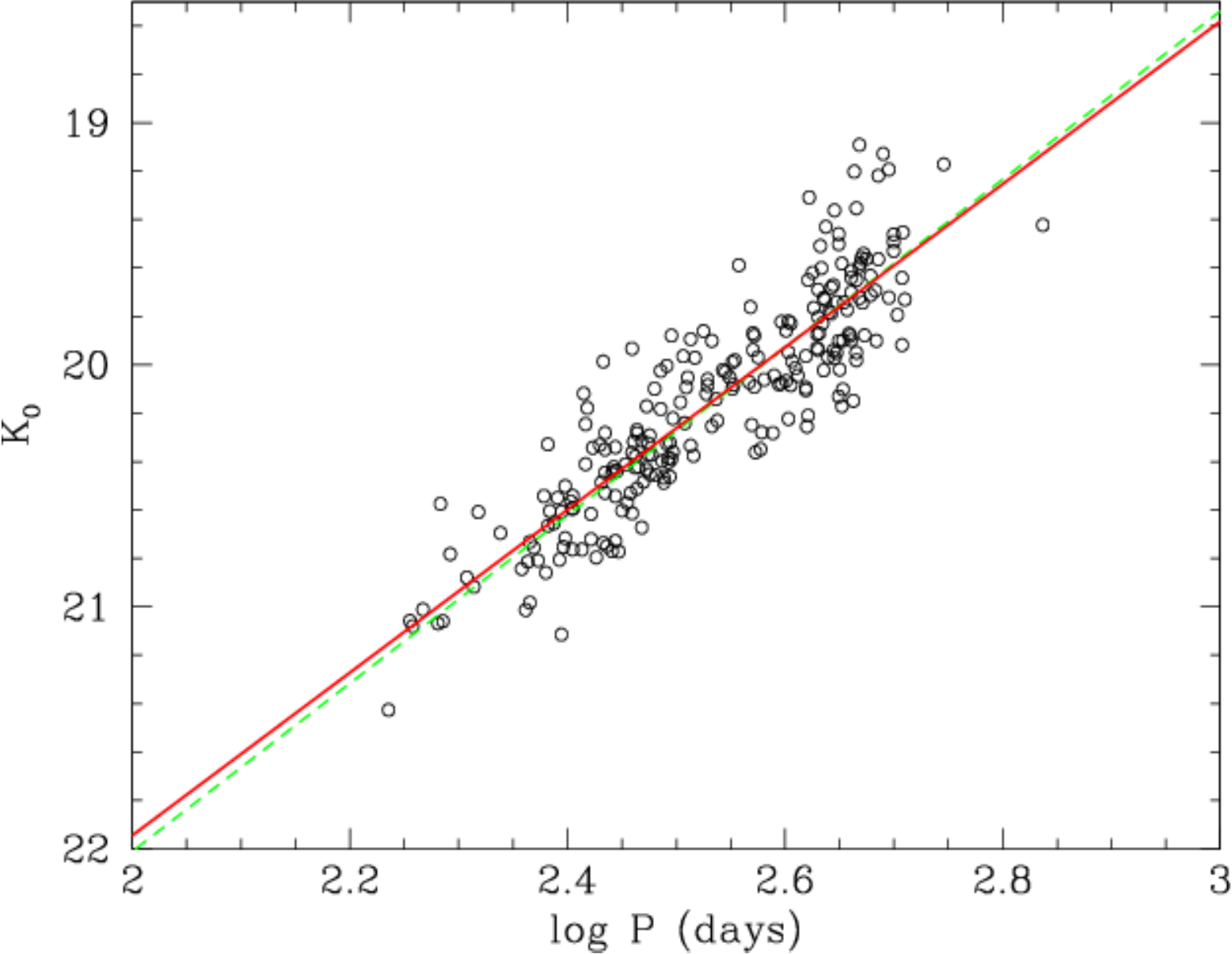}
\caption{Rejkuba (2004) derived a PLR for Miras with $J-K_S<1.4$ (presumed O-rich) in the inner halo of NGC\,5128. The red line is the best fit PLR while the green line is the LMC PLR (Feast {\it et al.} 1989).  }
 \label{Rejkuba}
\end{figure}

\subsection{Mira PLR at long periods} 
At periods over about 420 days most LMC O-rich stars are brighter than a linear PLR, fitted to the shorter period stars, would predict. Whitelock {\it et al.} (2003) suggested that this was due to hot bottom burning (HBB) and subsequent investigations (e.g. Menzies {\it et al.} 2015) support this interpretation. The additional energy provided by HBB presumably changes the structure of these stars and the characteristics of the PLR, although it is interesting to see that most LMC stars do still fall on a well defined PLR, albeit a different one from non-HBB stars. 

There is as yet no detailed understanding of HBB and  models differ one to another (e.g., Karakas {\it et al.} 2018, and references therein). It is not even clear what is the lowest mass at which HBB will occur, although there is consensus that it will depend on metallicity.  This is illustrated in the difference between the LMC PLR (e.g.,  Fig.~\ref{fig1}) and that of NGC\,5128 which was investigated by Rejkuba (2004) and is shown in Fig.~\ref{Rejkuba}.  The Miras in NGC\,5128 are in the inner halo which has a metallicity of  ${\rm [Fe/H]\sim -0.1}$, i.e. considerably higher than the LMC and probably has no HBB stars. They obey the same linear PLR as the short period LMC stars. 

Blommaert {\it et al.}  (2018) discuss OH/IR stars near the Galactic centre which appear to fall below the bolometric PLR, which they suggest is due to their being more evolved along the AGB than the bulk of Miras. This is an interesting finding that might be expected given the predictions of stellar evolution theory, although, as discussed above, a detailed understanding of the pulsation of stars in the fundamental mode remains elusive.  It is important that the existence of Miras with luminosities below the PLR is tested using the same method to measure the bolometric magnitude as was used to establish the calibrating magnitudes for the PLR (ideally using model fits to mean luminosities) and if at all possible in circumstances where interstellar reddening does not add significantly to the uncertainties. 

Possibly the most important point to make is that the PLR at $P>400$ days is going to differ in different galaxies, depending on the metallicity and mass range of that particular population.   If the PLR is to be used for distance determination then a linear version should be used and the analysis limited to stars with short periods. Long period Miras are discussed in more detail below. Because they are luminous and generally strong infrared sources they certainly have potential as distance scale probes. However, their behaviour is not yet understood sufficiently well that we can rely on them. 

\subsection{Most Distant Miras}
\begin{figure}
\centering
\includegraphics[height=6cm]{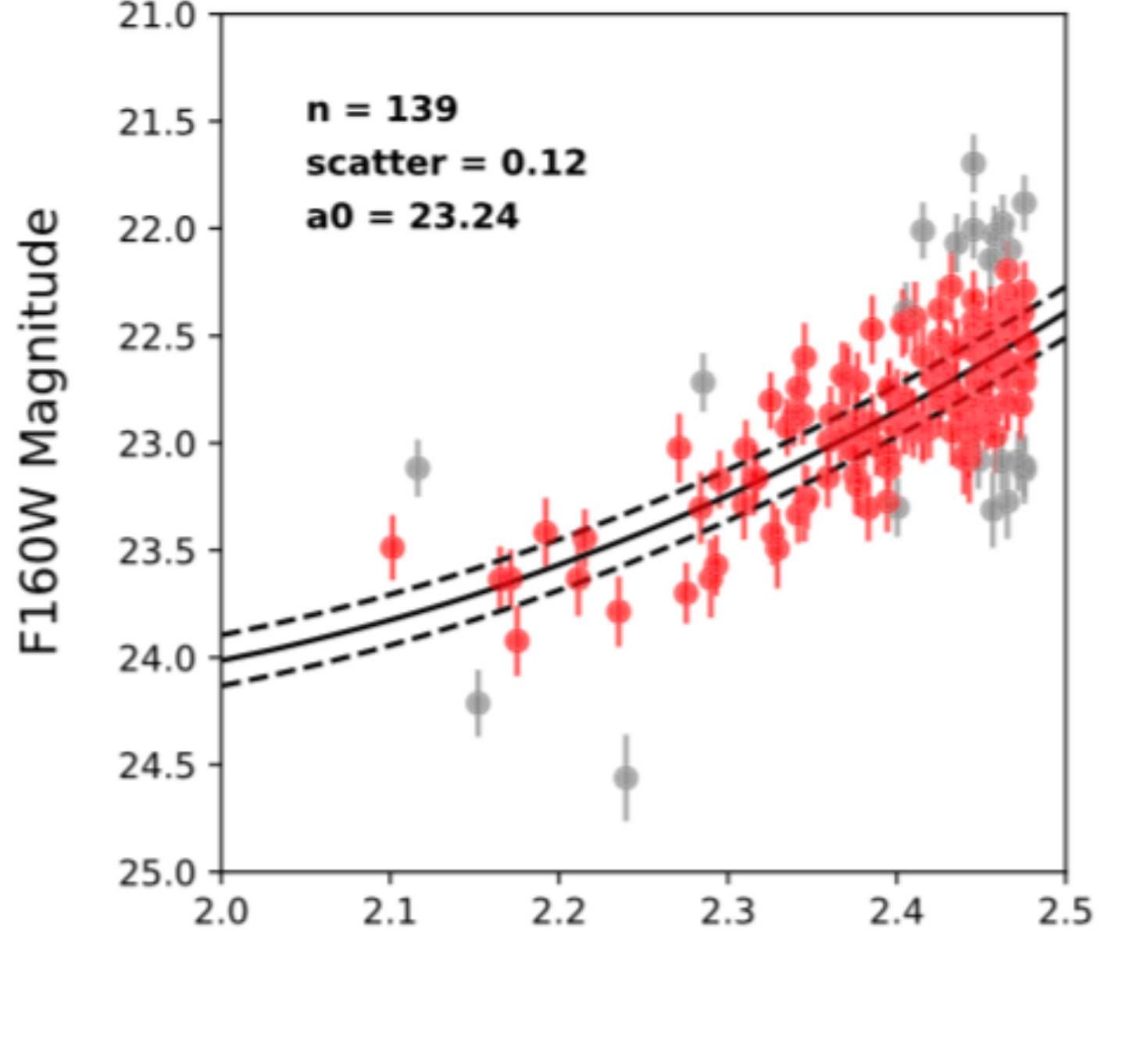}
\caption{PLR for the best quality sample of  Miras in NGC\,4258; the red points were those used to derive the final relation, while grey points were removed through 3$\sigma$ iterative clipping. The solid black curve shows the best fit relation and the dashed curves show the 1$\sigma$ fit. The functional form of the Yuan {\it et al.} (2017) fit was used and the zero-point was derived from the fit (Huang {\it et al. 2018). }}
 \label{Huang}
\end{figure}

Huang {\it et al.} (2018) used HST WFC3 observations to find Miras in NGC\,4258. This galaxy  was selected because it hosts a water megamaser and therefore has a well established distance (Riess {\it et al.} 2016). Observations of a single field were obtained, through the F125W and F160W filters, at 12 epochs spread over one year. 438 Mira candidates were identified of which 139 fitted the most stringent selection criteria and were used to measure the distance; these all had $P<300$ days and are illustrated in Fig.~\ref{Huang}.  Distances were measured relative to the LMC and agreed well with the Cepheids and the megamaser. At 7.5Mpc these are the most distant Miras to have measured periods and luminosities. This not only shows the potential of Miras to contribute to the distance scale problem, but opens possibilities for studying individual AGB variables to much larger distances with JWST.

\section{Miras as Probes of Stellar Evolution}
In this section I consider some recent work on Miras with long periods, $P>400$ days. Most of these have intermediate mass progenitors and the group includes hot bottom burning Miras, and stars that have been identified as `extreme AGB-stars', because of their thick circumstellar shells and the resultant very red colours. Many of them are C-rich, but some of the most interesting are O-rich and include OH/IR stars with periods over 1000 days. 

 HBB Miras in NGC\,6822, IC\,1613 and WLM have been discussed by Whitelock {\it et al.} (2013), Menzies {\it et al.} (2015) and Menzies (2018), respectively. Of particular interest are the very bright long period Miras discovered in the metal deficient Local Group galaxies, Sgr dIG and NGC\,3109 which  are discussed below.  It is also worth noting that there are very long period Miras in the SMC and LMC that were once thought to be supergiants, but which are almost certainly Miras; HV 11417 with P=1092, is one such example. Whether they are undergoing HBB is not clear. 

\begin{figure}
\centering
\includegraphics[height=9cm]{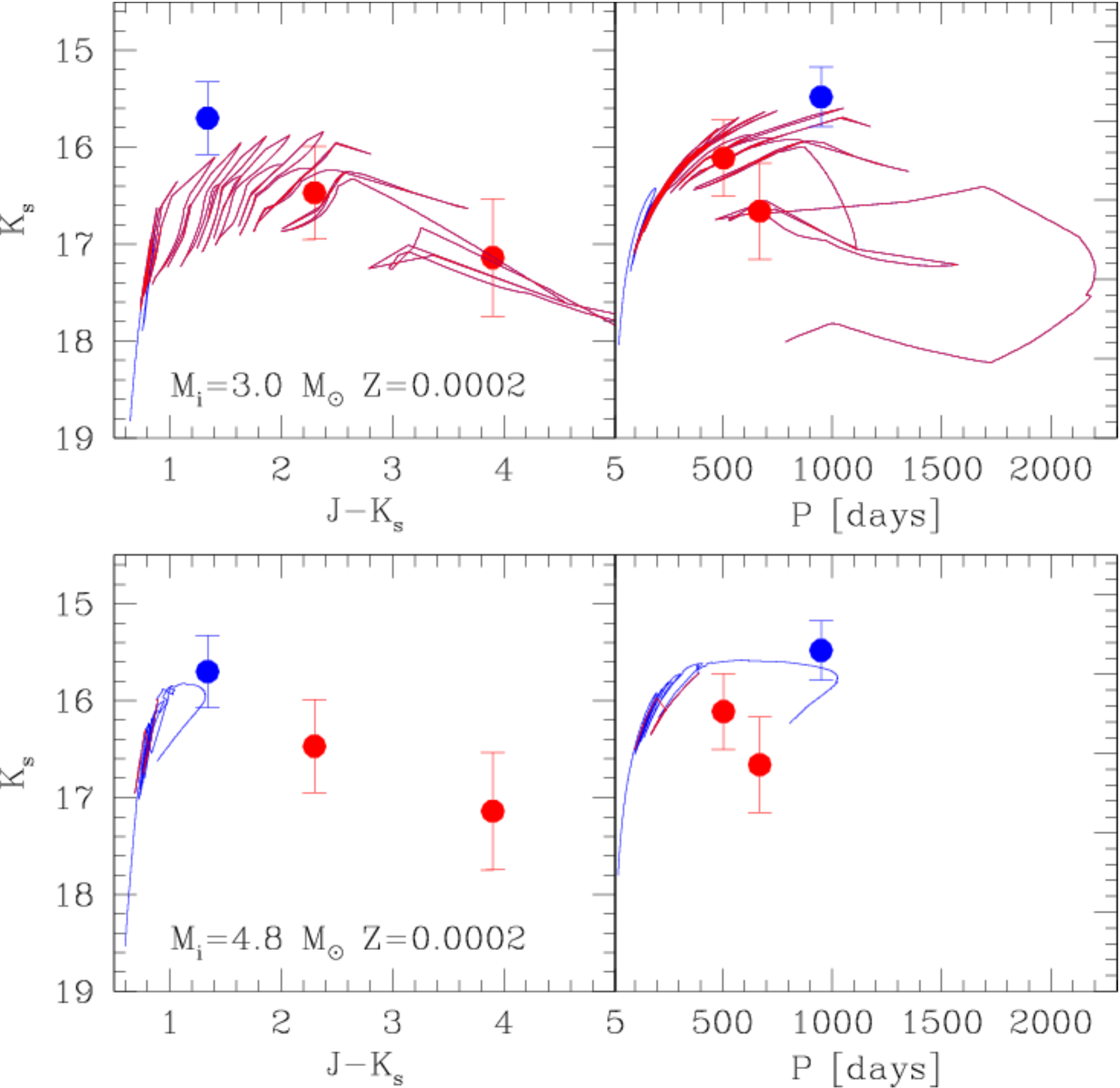}
\caption{Evolutionary tracks in  colour-magnitude and period-luminosity diagrams
(C stars in red and the O-rich star in blue).
Error bars show the variability range. AGB evolutionary tracks are shown for two choices of the initial mass as indicated and metallicity Z=0.0002. Stages characterized by surface $\rm C/O<1$  and $\rm C/O>1$ are coloured in blue and red, respectively (from Whitelock {\it et al.} 2018). } 
\label{tracks}
\end{figure}

\subsection{Miras in Sgr dIG}
Sgr dIG is a relatively low mass, Local Group, dwarf irregular that was surveyed at $JHK_S$ for variable stars, using the 1.4m InfraRed Survey Facility  (IRSF) by Whitelock {\it et al.} (2018). Three AGB variables were identified, two C-stars with periods of 504 and 670 days and one surprisingly blue ($(J-K_S)_0\sim 1.3$) O-rich Mira with P=950 days. At a distance of $(m-M)_0=25.2$ the IRSF is sensitive only to the brightest variables, so we can be certain that there are many more fainter Miras in this galaxy. The two C-rich Miras are similar to those found in other dwarf irregulars and a comparison with models from Marigo {\it et al.} (2017) suggests initial masses $M_i \sim 3 M_{\odot}$. The O-rich star is unusual; it has $M_{bol}\sim -6.7$ and a comparison with the models suggests an initial mass $M_i\sim 5 M_{\odot}$ and that it is in a short lived phase at the end of hot bottom burning.  Fig.~\ref{tracks} illustrates evolutionary tracks from Marigo {\it et al.} (2017) for both the C- and O-rich Miras.

\begin{figure}
\centering
\includegraphics[height=9cm]{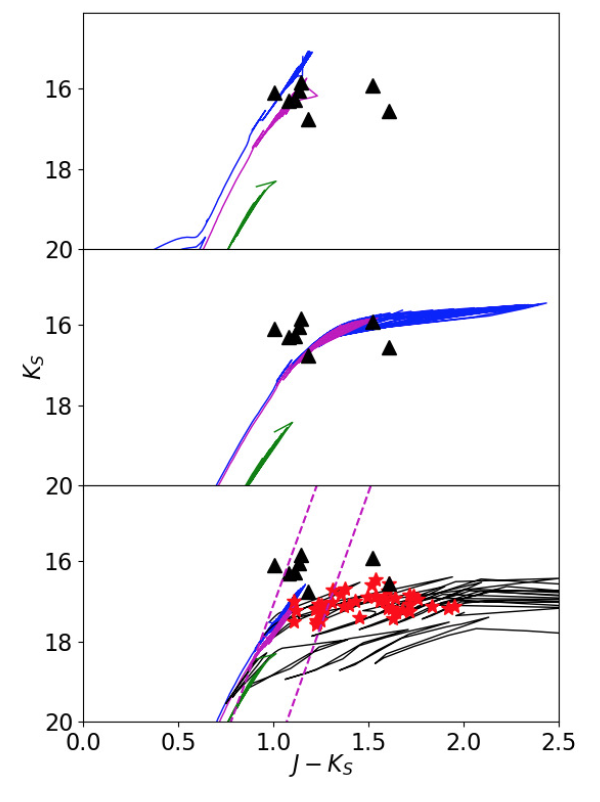}
\caption{Comparison of the NGC\,3109 AGB variables (black triangles) from Menzies {\it et al.} 2018 with Padova isochrones  (Marigo {\it et al.} 2017). (top) Isochrones for Z = 0.001 and ages of 0.1, 0.158 and 1 Gyr (later phases omitted). (middle) Isochrones for Z = 0.003 and ages of 0.158, 0.178 and 1 Gyr (later phases omitted). (bottom) Isochrones for Z = 0.001 and ages of 0.398, 0.501 and 1 Gyr and non-Mira C-stars illustrated as red stars. In the bottom plot, the black lines show the evolutionary phases where the models predict stars to have C/O$>$1; note agreement with the observations. The parallel dashed lines show the region where according to Cioni {\it et al.} (2006) only O-rich stars with Z = 0.001 should be found (it obviously does not apply to this galaxy). }

\label{N3109}
\end{figure}

\subsection{Miras in NGC\,3109}
NGC\,3109 is probably just outside the Local Group, has a low metallicity and is more massive than Sgr dIG. The IRSF $JHK_S$ survey was shallow and will therefore detect only the most luminous of the presumably numerous Miras. Menzies {\it et al.} (2018) found eight Miras, seven probably O-rich and one probably C-rich. Five of the O-rich candidates are very similar to the HBB stars found in IC\,1613, NGC\, 6822 and WLM.  These are the brightest of the blue ($J-K_S<1.2$) stars shown in Fig.~\ref{N3109}, where the isochrones in the top panel suggest ages of around 100 to 160 Myr. The other two O-rich candidates are only slightly older, the isochrone in the central panel of Fig.~\ref{N3109} suggests an age of around 180 Myr. The reddest of these has a period of 1486 days. The C star candidate Mira has a period of 1109 days and an age around 500 Myr.  If  its C-rich nature is confirmed will be amongst the longest period C-Miras known.  Miras with $P>1000$ days are unusual in the Galaxy and in the Magellanic Clouds and these are among the most massive and or most evolved AGB variables known as is discussed further by  Menzies {\it et al.} (2018).

\subsection{Miras in other galaxies}
The SPIRITS (Kasliwal {\it et al.} 2017; Karambelkar {\it et al.} 2018)  and DUSTiNGS collaborations (Boyer {\it et al.} 2015; Boyer 2018; Goldman {\it et al.} 2018) have together obtained multiple observations of large numbers of galaxies with the Spitzer spacecraft at 3.6 and 4.5 $\mu$m. They reveal numerous red, large amplitude, long period variables most of which will be Miras. We can anticipate learning a great deal about mass-loss and AGB evolution as we start to understand these stars.  

\section{Conclusions} 
Short period Miras are showing great potential as distance indicators, and their importance is likely to increase as accurate measurements at mid-infrared wavelengths become commonplace. Long period Miras, particularly those with $P>1000$ days are intriguing and there is need for more observations in different environments to understand the effects of mass and metellicity. There is also a great need for better understanding of pulsation and of evolution of these unusual stars. 

\acknowledgments I am grateful to my collaborators particularly John Menzies and Michael Feast for their part in this work and to John Menzies for a critical reading of a draft of this paper. My thanks to the South African National Research Foundation for a research grant.



\end{document}